\documentclass[aps, twocolumn, prl]{revtex4-1}
\usepackage{graphicx}
\usepackage{psfrag}
\usepackage{color}
\usepackage{latexsym}

\def\d {{\rm d}}

\begin{document}

\title{Universal, additive effect of temperature on the rheology of amorphous solids}
\author{Joyjit Chattoraj$^{(1)}$}
\author{Christiane Caroli$^{(2)}$}
\author{Ana\"el Lema\^{\i}tre$^{(1)}$}
\affiliation{$^{(1)}$ 
Universit\'e Paris Est -- Laboratoire Navier, ENPC-Paris, LCPC, CNRS UMR 8205
2 all\'ee Kepler, 77420 Champs-sur-Marne, France}
\affiliation{$^{(2)}$ INSP, Universit\'e Pierre et Marie Curie-Paris 6, 
CNRS, UMR 7588, 140 rue de Lourmel, 75015 Paris, France}

\date{\today}

\begin{abstract}
Extensive measurements of macroscopic stress in a 2D LJ glass, over a broad range of temperatures ($T$) and strain rates ($\dot\gamma$), demonstrate a very significant decrease of the flowing stress with $T$, even much below the glass transition. A detailed analysis of the interplay between loading, thermal activation, and mechanical noise leads us to propose that over a broad ($\dot\gamma,T$) region, the effect of temperature amounts to a mere lowering of the strains at which plastic events occur, while the athermal avalanche dynamics remains essentially unperturbed. Temperature is then shown to correct the athermal stress by a (negative) additive contribution which presents a universal form, thus bringing support to and extending an expression proposed by Johnson and Samwer~\cite{JohnsonSamwer2005}. Our prediction is shown to match strikingly well numerical data up to the vicinity of $T_g$.
\end{abstract}

\maketitle

It is now well accepted that macroscopic plastic deformation of amorphous solids is the net result of an accumulation of local rearrangements (``shear transformations'', or simply ``flips'') involving small clusters of atoms or particles~\cite{Argon1979}. However, what controls the occurrence of these flips, ie when and where they take place, remains a controversial question, which needs to be answered prior to the formulation of any theory of plasticity.

Recent numerical works bring fragmentary information about these mechanisms. 
In athermal simulations--whether quasi-static or at finite strain-rate--it is possible to identify, at any time, in the steady-state sheared system a set of soft zones, which retain their identity for sizeable stretches of strain~\cite{LemaitreCaroli2007}. As a zone is loaded by the external drive, it gradually softens until it reaches its instability threshold and flips~\cite{MaloneyLemaitre2004a}. This event creates in the surrounding medium a long-ranged elastic field, with quadrupolar symmetry~\cite{Eshelby1957}, which shifts the strain of other zones, hence may induce secondary events. This mechanism gives rise to avalanches of flips~\cite{MaloneyLemaitre2004,*BaileySchiotzLemaitreJacobsen2007} with a strain-rate dependent average size~\cite{LemaitreCaroli2009}. 

These results have led us to propose a picture in which, at $T=0$, the dynamics of soft zones in the vicinity of their instability thresholds plays a critical role in controlling dissipative events, while structural disorder, which permits the existence of soft zones in the first place, is otherwise accessory to the unfolding of plastic events. But we must then ask what is the effect of a finite temperature on plastic activity~\cite{HentschelKarmakarLernerProcaccia2010}. Indeed, recent works concerned with stress and elastic fluctuations, as well as energy barrier distributions, have shown that strained systems under steady flow present a sizeable fraction of low-energy barriers, most of which are not involved in the athermal response~\cite{RodneySchuh2009a}. It is then possible that, even a small temperature may activate jumps over these many other barriers, hence disrupting the avalanche processes which control the athermal response.

%Here, we analyze in detail the effect of thermal noise on the dynamics of zones driven towards the saddle-node bifurcations corresponding to their instability thresholds. We show that the competition between loading and thermal activation results in the definition of effective thresholds which are separated from the mechanical yield points by a $\dot\gamma$- and $T$-dependent shift. We then show that in a broad region of the $(\dot\gamma,T)$ parameter space.

Here, we perform extensive measurements of the macroscopic stress in a 2D Lennard-Jones glass, over a broad range of temperatures ($T$) and strain rates ($\dot\gamma$). We find that finite temperatures induce a very significant decrease of the flowing stress. We show that these data can be interpreted within the framework provided by a detailed analysis of the interplay between loading, thermal activation, and mechanical noise. This analysis predicts that over a broad region of the $(\dot\gamma,T)$ parameter space, the avalanche dynamics should remain basically unperturbed, the effect of temperature amounting to a mere lowering of the strains at which plastic events occur, leading to an expression for stress of the form:
$$
\sigma(\dot\gamma,T) = \sigma_0(\dot\gamma) + \Phi(\dot\gamma,T)
$$
where (i) $\sigma_0(\dot\gamma)$ exhibits precisely the dependence found in athermal systems and (ii) the (negative) shift function $\Phi\propto-(-T\ln(C\,\dot\gamma/T^{5/6}))^{2/3}$ is analogous to that proposed by Johnson and Samwer~\cite{JohnsonSamwer2005} to account for a number of experimental results on metallic glasses. This expression matches strikingly well our numerical data up to the vicinity of the glass transition temperature $T_g$. We thus conclude to the robustness of avalanches, which also show up clearly in the same temperature range in the structure of strain maps.
%This, we argue, entails that in this broad range of temperatures, the unfolding of avalanches is essentially unperturbed, while the effect of temperature reduces to a downward shift of the strain at which plastic are triggered, due to the competition between thermal activation and zone softening.
%This, we argue, entails that the effect of temperature amounts to a downward shift of the strain at which plastic events are triggered, which results from the competition between thermal activation and zone softening, the unfolding of avalanches being otherwise unchanged. 
%We thus conclude to the robustness of avalanches, which also show up clearly in the structure of strain maps, up to the vicinity of the glass transition.

We use the same 2D binary LJ mixture as in Refs.~\cite{LemaitreCaroli2007,LemaitreCaroli2009}: it is composed of large~(L) and small~(S) particles with radii $R_L=0.5, R_S=0.3$ and equal masses $m=1$ (in standard LJ units), in a number ratio of $N_L/N_S=(1+\sqrt{5})/4$ to ensure that no crystallization occurs, at packing fraction $\pi(N_LR_L^2+N_SR_S^2)/L^2=0.9$. With these parameters the shear modulus is $\sim20$~\cite{LemaitreCaroli2007}, the shear wave speed $c_s\simeq3.4$, and the glass transition temperature (identified as the value at which $\tau_\alpha=10^4$, and computed following~\cite{PereraHarrowell1999}) is $\sim0.27$. Finite temperature MD simulations are performed on square $L\times L$ systems, using Lees-Edwards boundary conditions, with a standard velocity rescaling protocol~\cite{AllenTildesley1996} and time steps $dt\leq0.01$. All the data presented here are obtained after $100\%$ of strain, which ensures that the system is in steady state. Large sets and long strain intervals are used for statistical accuracy (e.g. for $L=40$, $25$ samples are strained up to 1300\%).

\begin{figure}[h]
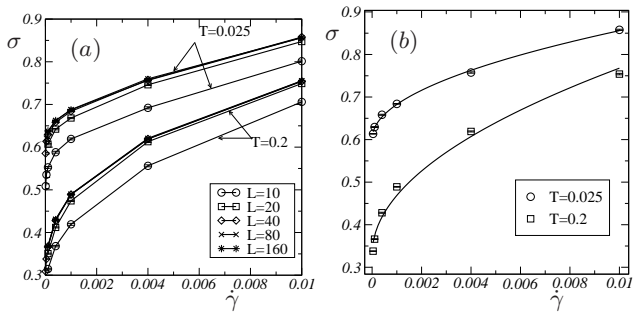

\psfrag{a}{{$(a)$}}
\psfrag{b}{{$(b)$}}
\psfrag{s}{{$\sigma$}}
\psfrag{g}{{$\dot\gamma$}}
\psfrag{t}{{$T$}}
\psfrag{tmix}{{$T\searrow$}}
\psfrag{gmix}{{$\dot\gamma\searrow$}}
\includegraphics*[width=0.23\textwidth]{sigma-gdot-collapse.eps}
\includegraphics*[width=0.23\textwidth]{L-40-sigma-gdot-athermalfit.eps}
\caption{
(a): The macroscopic stress $\sigma$ as a function of strain rate $\dot\gamma$ 
for systems of sizes $L=10$, 20, 40, 80, and 160, at temperatures $T=0.025$ and $T=0.2$.
(b): $\sigma$ versus $\dot\gamma$ for $L=40$, compared with 
fits of the form $\sigma=A_0+A_1\,\sqrt{\dot\gamma}$ (solid lines).
}
\label{fig:stress}
\end{figure}

We present on Fig.~\ref{fig:stress}-(a) steady state stress $\sigma$ vs strain rate $\dot\gamma$ for different system sizes $L=10$, 20, 40, 80, 160, two temperatures $T=0.025$ and $T=0.2$, and $\dot\gamma$ ranging from $4\times10^{-5}$ to $10^{-2}$. For each temperature we find, as in athermal systems, quick convergence with increasing $L$, of $\sigma(\dot\gamma)$ towards a master curve. In the range of $\dot\gamma$ studied, saturation is already reached for $L=40$, which allows us to focus in the following on stress data obtained for this system size.

To qualify the effect of temperature, we then attempt (see Fig.~\ref{fig:stress}-(b)) to fit these curves with $\sigma=A_0+A_1\,\sqrt{\dot\gamma}$, a form which was shown to match very well the results of athermal simulations. For our lowest temperature $T=0.025$, the fit is quite satisfactory, although not as good as was found with athermal data. This suggests that at $T$ finite but $\ll T_g$ the dynamics is closely similar to that of an athermal system. At higher temperature, however, not only does this fit become definitely poorer but, more importantly, the overall stress level drops significantly: temperature thus has a pronounced effect on dissipation. Does this mean that it completely modifies the avalanche-dominated dynamics at work in the $T\to0$ limit?
% mechanisms at work

We know that plastic events correspond to local shear transformations (zone flips), which are preceded by the gradual vanishing of an energy barrier under loading. At $T=0$, each flip is triggered right at the threshold strain $\gamma_c$ where instability is reached. At finite $T$, we must expect that flips occur in anticipation due to thermal activation, which becomes gradually more efficient when a zone approaches its $\gamma_c$~\cite{CaroliNozieres1996}. More precisely, consider a single zone, initially lying at a strain $\gamma_0<\gamma_c$, which is loaded at finite strain rate $\dot\gamma$ in the presence of thermal noise. At an external strain $\gamma\in[\gamma_0,\gamma_c[$, the probability $P(\gamma)$ that it has not yet flipped obeys $P(\gamma+\d\gamma) = P(\gamma)\left(1-\frac{\d\gamma}{\dot\gamma}\,R(\gamma)\right)$, or:
$$
\frac{\partial P}{\partial\gamma}=-\frac{1}{\dot\gamma}\,R(\gamma)\,P(\gamma;\gamma_0)
$$
with $R(\gamma)$ the rate of activated jumps.
The solution is:
\begin{equation}
\label{eqn:p}
P(\gamma;\gamma_0) = \exp\left[-\frac{1}{\dot\gamma}\int_{\gamma_0}^\gamma\d\gamma' R(\gamma')\right]
\end{equation}
At low temperature, activation is efficient only close to the saddle node bifurcation occurring at $\gamma_c$. There, the energy barrier and the ``attempt frequency'' present the universal scaling forms: $\Delta E=B\,(\gamma_c-\gamma)^{3/2}$ and $\omega=\nu\,(\gamma_c-\gamma)^{1/4}$~\cite{MaloneyLemaitre2004a}\footnote{The vanishing eigenvalue controlling the saddlenode bifurcation scales as $\lambda\propto(\gamma_c-\gamma)^{1/2}\propto\omega^2$.}. Provided that $\Delta E/T\gg 1$, we can use the standard Kramers expression for the activation rate: $R(\gamma)=\omega\,\exp(-\Delta E/T)$, so that:
$$
P(\gamma;\gamma_0) = \exp\left(-\frac{2}{3}\,\frac{\nu}{\dot\gamma}\,\left(\frac{T}{B}\right)^{5/6}\,\Big[Q(\delta\gamma)-Q(\delta\gamma_0)\Big]\right)
$$
where $\delta\gamma=\gamma_c-\gamma$ and $Q(\delta\gamma) = \Gamma\left(\frac{5}{6}; \frac{B}{T}\,\delta\gamma^{3/2}\right)$,
with $\Gamma$ the upper incomplete gamma function. Our usage of Kramers' expression implies that $\epsilon={T}/B{\delta\gamma}^{3/2}\ll 1$, so that we can use the asymptotic expression $\Gamma(s;x)\sim x^{s-1}\,e^{-x}$, whence $Q(\delta\gamma)\sim \epsilon^{1/6}\,\exp(-1/\epsilon)$.

\begin{figure}[h]
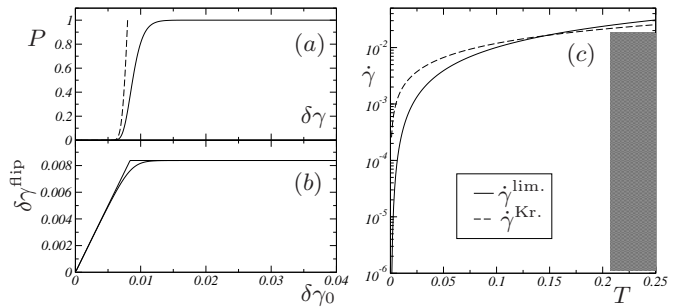

\psfrag{gkr}[][l]{{$\,\,\,\,\dot\gamma^{\rm Kr.}$}}
\psfrag{glm}[][l]{{$\,\,\,\,\dot\gamma^{\rm lim.}$}}
\psfrag{g0}{{$\delta\gamma_0$}}
\psfrag{dg}{{$\delta\gamma$}}
\psfrag{dgf}{{$\delta\gamma^{\rm flip}$}}
\psfrag{g}{{$\dot\gamma$}}
\psfrag{p}{{$P$}}
\psfrag{T}{{$T$}}
\psfrag{a}{{$(a)$}}
\psfrag{b}{{$(b)$}}
\psfrag{c}{{$(c)$}}
\includegraphics*[height=0.22\textwidth]{p.eps}
\hfil
\includegraphics*[height=0.22\textwidth]{domain.eps}
\caption{
The parameters used come from from the fit of our stress data (see text). For $T=0.1$: (a) the function $P$ vs $\delta\gamma$ for $\gamma_0=-\infty$ (solid line) and $\gamma_0\gtrsim\gamma^\star$ (dashed); (b) $\delta\gamma^{\rm flip}$ vs $\delta\gamma_0$ (solid line) with asymptotes (thin line). (c): $\dot\gamma^{\rm lim.}$ and $\dot\gamma^{\rm Kr.}$ vs $T$ (using $a/d=1/3$, $\Delta\epsilon_0=5\%$); shaded area: the high-$T$ region where data start to depart from fit. %in the $\dot\gamma$ vs $T$  plane.
}
\label{fig:plots}
\end{figure}
Plots of $P(\gamma;\gamma_0)$ are presented on Fig.~\ref{fig:plots}-(a). Let us first consider the formal limit $\gamma_0\to-\infty$ ($Q(\delta\gamma_0)\to0$), which corresponds to the case when a zone is initially very far below threshold. Because $P$ is an exponential of an exponential it presents a very sharp transition from $P\sim1$ to $P\sim0$ around a strain $\gamma^\star$ such that:
\begin{equation}
\label{eqn:cond}
\frac{2}{3}\frac{\nu}{\dot\gamma}\,\left(\frac{T}{B}\right)^{5/6}\,Q(\delta\gamma^\star) = 1
\quad,
\end{equation}
%\begin{equation}
%\label{eqn:cond}
%\frac{\nu}{\dot\gamma}\,e^{-B\,\frac{{\delta\gamma^\star}^{3/2}}{k_B\,T}}\,\left[\frac{3}{2}\,\frac{{\delta\gamma^\star}^{3/2}}{k_B\,T}+\frac{1}{4}\right]^{-1}\,{\delta\gamma^\star}^{5/4} = 1
%\end{equation}
%We see here that the effect of a finite temperature is to trigger plastic events at an apparent threshold $\gamma_c-\delta\gamma^\star$ which is shifted down from $\gamma_c$. An estimate for $\delta\gamma^\star$ is next obtained by expanding equation~(\ref{eqn:cond}) to leading order in $\epsilon={k_B\,T}/A{\delta\gamma^\star}^{3/2}$:
which is next solved at leading order in $\epsilon$, yielding:
\begin{equation}
\label{eqn:shift}
\delta\gamma^\star\simeq \left[\frac{T}{B}\ln\left(\frac{2}{3}\frac{\nu}{\dot\gamma}\,\left(\frac{T}{B}\right)^{5/6}\right)\,\right]^{2/3}
\quad.
\end{equation}
The width of this transition is of order $|\partial P/\partial\gamma (\gamma^\star;-\infty)|^{-1}=({2e}/{3})\,\epsilon^\star\,\delta\gamma^\star$, which is thus $\mathcal{O}(\epsilon^\star)$ relative to $\delta\gamma^\star$ itself.

As $\gamma_0$ increases from $-\infty$, the curves $P(\gamma;\gamma_0)$ remain nearly identical to $P(\gamma;-\infty)$ up to the immediate vicinity of $\gamma^\star$. For $\gamma_0>\gamma^\star$, $P(\gamma;\gamma_0)$ no longer presents a plateau at low values of $\gamma$, but drops already sharply at $\gamma_0$, with slope $|\partial P/\partial\gamma (\gamma_0;\gamma_0)|=\frac{1}{\dot\gamma}\,R(\gamma_0)$, which is an increasing function of $\gamma_0$. Therefore, the solution $\gamma^{\rm flip}$ of $P(\gamma^{\rm flip};\gamma_0)=1/e$ can always be interpreted as the typical strain at which a zone starting from $\gamma_0$ flips.

The curve $\gamma^{\rm flip}(\gamma_0)$ (Fig.~\ref{fig:plots}-(b)) exhibits a sharp transition between two limiting behaviors: (i) when injected at $\gamma_0\lesssim\gamma^\star$, a zone flips at $\gamma^{\rm flip}\approx\gamma^\star$; (ii) when injected at $\gamma_0\gtrsim\gamma^\star$, it flips almost immediately: $\gamma^{\rm flip}\approx\gamma_0$. The width of the transition can be estimated as being $\mathcal{O}(\delta\gamma^\star\,\epsilon^\star)\ll \delta\gamma^\star$, which compares with the width of the drop of $P(\gamma;-\infty)$ around $\gamma^\star$. We can therefore conclude that the competition between thermal activation and drive defines (slightly fuzzy) apparent thresholds which are shifted by $-\delta\gamma^\star(\dot\gamma,T)$ from the mechanical yield points.

In our previous studies of athermal systems, we found that the mechanical noise--that is the stress noise generated by the flips themselves--played a key role by inducing correlations between flip events, leading to the emergence of avalanche behavior. At finite $T$, a zone embedded in a sheared system is thus experiencing both thermal noise and random strain shifts (ie barrier height fluctuations) originating from prior events. In~\cite{LemaitreCaroli2009}, we proposed to separate mechanical noise into: (i) a low-frequency part, generated by nearby events (within a sphere of radius $\ell$); (ii) a background noise coming from the rest of the system. Requiring that the near-field signals be non-overlapping and stand out of the background noise defines a single length $\ell(\dot\gamma)$ which we identified as the avalanche size.

An exact treatment of the jump dynamics in the presence of activation, loading, and barrier fluctuations is for the moment out of reach. However we note that the values $\gamma^\star$ can still be interpreted as effective thresholds, provided that the unfolding of the activated jump of a zone near its $\gamma^{\rm flip}$ is not perturbed by the ambient noise (which includes nearby and far-field signals). For this purpose, we recall that the strain field due to a shear transformation has the form $f_{\rm esh.}=\frac{a^2}{r^2}\Delta\epsilon_0$~\cite{Eshelby1957}, where $a$ is a zone size and $\Delta\epsilon_0$ an elementary plastic strain. In a system of size $L$, the flip rate is ${\mathcal R}=\dot\gamma L^2/(a^2\Delta\epsilon_0)$, and the strain fluctuation due to the noise incoming from the whole system during a time $\tau$ verifies $\Delta\gamma^2(\tau)=\frac{\tau\,\mathcal{R}}{L^2}\,\int_{d}^L f_{\rm esh.}^2\d^2 r=\dot\gamma\,\tau\,\frac{a^2\,\Delta\epsilon_0}{d^2}$, with $d$ a typical inter-zone distance~\cite{LemaitreCaroli2009}. In order for the activation process to be negligibly perturbed, we must ensure that the strain fluctuation accumulated during the activation time $1/R(\gamma^{\rm flip})<1/R(\gamma^\star)$, remains much smaller than the thermal strain shift $\delta\gamma^\star$ itself. This is guaranteed as soon as: $\dot\gamma\,{a^2}\Delta\epsilon_0/(R(\gamma^\star)d^2)\ll (\delta\gamma^\star)^2$. As $R(\gamma^\star)=\frac{3}{2}\frac{B}{T}\,\dot\gamma\,\sqrt{\delta\gamma^\star}$, this also writes:
\begin{equation}
\label{eqn:condition}
%\delta\gamma^\star\gg\left(\frac{2}{3}\,\frac{T}{B}\,\frac{a^2\Delta\epsilon_0}{d^2}\right)^{2/5}
\dot\gamma\ll\dot\gamma^{\rm lim.}=\frac{2\nu}{3}\bigg(\frac{T}{B}\bigg)^{5/6}e^{-\big(\frac{B}{T}\big)^{2/5}\big(\frac{2}{3}\frac{a^2\Delta\epsilon_0}{d^2}\big)^{3/5}}
\end{equation}

When this condition is satisfied, the process of flip activation disentangles from the response to incoming mechanical noise signals. The basic elements of a phenomenology of avalanche dynamics, namely the advection of zones towards (shifted) thresholds and the presence of mechanical noise signals, are preserved. In particular, the separation between nearby, correlated, signal and background noise proceeds exactly along the same lines as in the $T=0$ limit~\cite{LemaitreCaroli2009}, thus defining the same value for the avalanche size $\ell(\dot\gamma)$. As flip events occur at shifted thresholds, the macroscopic stress should thus be of the form:
\begin{equation}
\label{eqn:sigma}
\sigma(\dot\gamma, T) = \sigma_0(\dot\gamma)-\mu\,\overline{\delta\gamma^\star}(\dot\gamma, T)
\end{equation}
where $\mu$ is the shear modulus. From athermal simulations~\cite{LemaitreCaroli2009}, we know that $\sigma_0$ is of the form: $\sigma_0(\dot\gamma)=A_0+A_1\,\sqrt{\dot\gamma}$. The average $\overline{\delta\gamma^\star}$ accounts for the fact that the values of $B$ and $\nu$ appearing in the above calculation are distributed due to structural disorder. Assuming that $\ln B$ and $\ln \nu$ are well-centered, we can finaly write:
\begin{equation}
\label{eqn:fit}
\sigma(\dot\gamma) = A_0+A_1\,\sqrt{\dot\gamma} - A_2\,T^{2/3}\,\left[\ln\left(A_3\,T^{5/6}/\dot\gamma\right)\right]^{2/3}
\end{equation}
with $A_2=\mu \overline{B^{-2/3}}$ and $A_3 = \frac{2}{3}\overline{{\nu}/{B^{5/6}}}$. 

\begin{figure}[b]
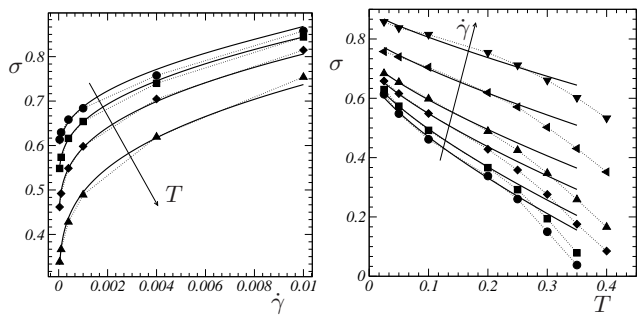

\psfrag{s}{{$\sigma$}}
\psfrag{g}{{$\dot\gamma$}}
\psfrag{T}{{$T$}}
\psfrag{tmix}{{$T\searrow$}}
\psfrag{gmix}{{$\dot\gamma\searrow$}}
\includegraphics*[width=0.23\textwidth]{L-40-stress-gdot3-fit.eps}\hfil
\includegraphics*[width=0.23\textwidth]{L-40-stress-T3-fit.eps}
%\hfil\includegraphics*[width=0.25\textwidth]{domain.eps}
\caption{
Macroscopic stress $\sigma$ (filled symbols) as a function of strain rate $\dot\gamma$ (left) and $T$ (right) compared with the fit obtained using Eq.~(\ref{eqn:fit}), with parameters: $A_0=0.66$, $A_1=2.09$, $A_2=0.27$, and $A_3=0.22$.
}
\label{fig:fit}
\end{figure}

To test this prediction, we now fit the stress data shown on Fig.~\ref{fig:stress} using this four parameter expression. As seen on Fig.~\ref{fig:fit}, the fit is remarkable over quite a broad range of parameters, $T$ varying from nearly 0 up to $\sim0.2$ (to be compared with $T_g=0.27$), and $\dot\gamma$ ranging over more than two decades. The values of the parameters are strongly constrained by the fit: $A_0$ and $A_1$ by the low-$T$ data; $A_2$ by the low-$\dot\gamma$ data. It clearly confirms the form of $\sigma_0(\dot\gamma)$ as well as the functional form of the correction term, except for the $5/6$ exponent which is weakly discriminated. The $T^{2/3}$ dependence, which is very visible at low $\dot\gamma$, is a clear signature that the saddle-node bifurcation controls the behavior of energy barriers heights over the whole range of relevants strains.
\begin{figure}[h]
\includegraphics[width=0.43\textwidth]{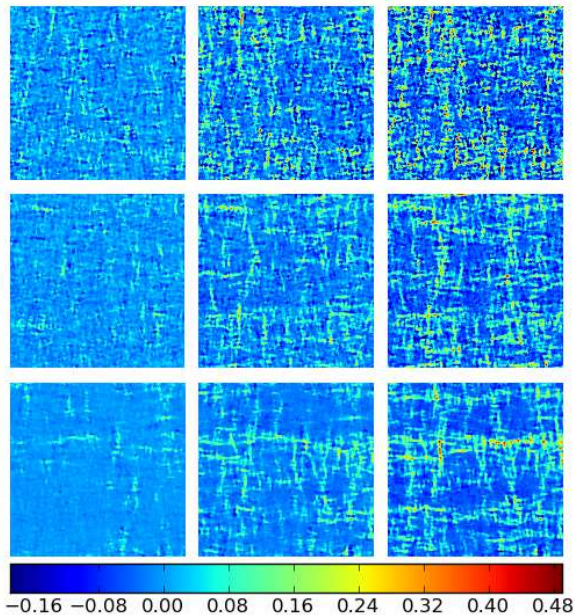}
\caption{
The non-affine strain field accumulated over strain intervals $\Delta\gamma=1\%, 5\%, 10\%$ (left to right),
when sheared at $\dot\gamma=10^{-4}$, and for $T=0.025,0.2,0.3$ (bottom to top). System size is $160\times160$.
}
\label{fig:strain}
\end{figure}

From the fits we extract typical values $\nu\sim50$ and $B\sim650$, from which we can examine a posteriori the validity of %XXX analysis XXX 
our expression for stress. The applicability of the Kramers expression ($1/\epsilon^{\star}=\ln(({2}{\nu}/3{\dot\gamma})\left({T}/{B})^{5/6}\right)\gg 1$) and condition~(\ref{eqn:condition}) both set upper limits $\dot\gamma^{\rm Kr.}$ and $\dot\gamma^{\rm lim.}$ (resp.), which are increasing functions of $T$ (see Fig.~\ref{fig:plots}-(c)). 
%Their values for a few temperatures are listed in the following table:
%\begin{center}
%\begin{tabular}{|l|c|c|c|c|} \hline
%$T$ & 0.025 & 0.05 & 0.1 & 0.2 \\ \hline
%$\dot\gamma^{\rm Kr.}$ & 2.5\times10^{-3} & 4.3\times10^{-3} & 7.5\times10^{-3} & 1.4\times10^{-2} \\ \hline
%$\dot\gamma^{\rm lim.}$ & 8\times10^{-4} & 2.5\times10^{-3} & 6.5\times10^{-3} & 1.5\times10^{-2} \\ \hline
%\end{tabular}
%\end{center}
By iteration, we have used for our fit only the points which satisfy these conditions. This excludes a few data points at the lowest temperatures and highest strain rates. 

The conditions $\dot\gamma<\dot\gamma^{\rm Kr.}(T)$, $\dot\gamma<\dot\gamma^{\rm lim.}(T)$ and $T\lesssim0.2$ define a large region of parameter space within which we claim that the effect of thermal fluctuations reduces to a universal additive contribution to stress of the form given in Eq.~(\ref{eqn:sigma}). In the vicinity of $T_g$, the gradual departure of the measured stress away from this expression should be assigned to the increasing contribution of thermally activated events of a different nature than the zone flips that we have considered here. We believe, these might be related to the finite, low-lying barriers shown by Rodney and Schuh~\cite{RodneySchuh2009a} to be present in sheared systems.

The present work provides a firm support to the notion of effective threshold and generalizes significantly the expression shown by Johnson and Samwer to fit a large body of experimental data on metallic glasses~\cite{JohnsonSamwer2005}. Indeed, our derivation clarifies why an additive correction to stress, with the universal form set by the saddle-node bifurcation, holds despite the presence of structural disorder. Moreover, our data show that this correction adds up, as predicted by our model, to a stress $\sigma_0(\dot\gamma)$ which is precisely that expected in an athermal system. We thus conclude to the robustness of avalanche dynamics up to the vicinity of $T_g$. 

In order to gain further insight into this question, we now show how the non-affine strain field accumulates, at different temperatures, as the system is macroscopically sheared. Strain maps are displayed on Fig.~\ref{fig:strain} for increasing values of the external strain $\Delta\gamma=1\%, 5\%, 10\%$ (left to right), and for $T=0.025,0.2,0.3$ (bottom to top). Strikingly enough, at all values of $T$, we clearly see comparable correlated structures. The strain patterns at $T=0.025$ and 0.2 are similar and exhibit, as those seen previously in athermal simulations, a clear directionality which can only be due to Eshelby-like elastic interactions~\cite{LemaitreCaroli2009}. It is only at $T=0.3$, i.e. in the supercooled regime, that a qualitative change can be observed, the structures becoming slightly blurred and shorter-ranged. We thus see a continuity of behavior across the glass transition which connects the low-temperature avalanche behavior with the cooperative dynamics observed by Tanaka in the supercooled regime~\cite{FurukawaKimSaitoTanaka2009}. More quantitative information on this question might come from the detailed analysis of the $(\dot\gamma,T)$-dependence of the diffusion coefficient, a study which is presently under way.

This work was supported by
the French competitiveness cluster Advancity and
Region \^Ile de France.

\end{document}